\documentclass[11pt]{article}
\usepackage{epsf}
\usepackage{graphicx}        
\def\ll{\label}
\def\re{\ref}
\def\c{\cite}

\def\r1{(\ref{$1})}

\def\ba{\begin{array}{c}}

\def\ea{\end{array}}

\def\l{\left}
\def\l({\left(}
\def\r){\right)}
\def\r{\right}

\def\la{\lambda}

 \def\be{\begin{equation}}
\def\bc{\begin{center}}
\def\ec{\end{center}}
\def\bit{\begin{itemize}}
\def\eit{\end{itemize}}
\def\ee{\end{equation}}
\def\ed{\end{document}}
\def\bea{\begin{eqnarray}}
\def\eea{\end{eqnarray}}
\def\efr{\end{flushright}}

\begin{document}
\title{Novel integrable higher-dimensional  nonlinear Schr\"odinger
equation: properties, solutions, applications}
\author{ Anjan Kundu$^1$, Abhik Mukherjee$^2$
\\
 Theory Division,
Saha Institute of Nuclear Physics\\
 Kolkata, INDIA\\$^1$
 anjan.kundu@saha.ac.in
\\ $^2$abhik.mukherjee@saha.ac.in
}
 \maketitle

\noindent

\begin{abstract} 
 An  integrable extension of the well known nonlinear Schr\"odinger (NLS)
equation to a higher space-dimension, recently proposed by us, is
investigated, exploring its various important aspects.
 Focusing  on  the idea of construction its connection with other known
models like the Zakharov equation and the Strachan construction is shown. 
The underlying integrable structures like the Lax pair, the infinite
conserved charges and the higher soliton solutions are presented in the
explicit form.  The related 2D rogue wave model and
   other  applications are focused on. 
 \end{abstract}
02.30.lk,
 05.45.Yv,
02.30.jr,
92.10.H+,
42.65.-k,
\smallskip

\noindent {\bf keywords}: { { Integrable 2D NLS equation, 
Lax pair, exact  solitons, 2D ocean rogue wave}
}
\smallskip

\noindent Short title : {\it Integrable 2D NLS  }


\section{Introduction:}

In the development of  nonlinear integrable systems, the $(1+1)
$-dimensional NLS equation, though being a rather late entrant \c{solit},
gained popularity very quickly due to its enormously rich theoretical and
practical importance.  Over the years, the 1D NLS model has
 penetrated into different subjects with its  soliton solutions   finding
important applications in diverse fields, starting from nonlinear plasma,
  optical
 communication to
  nonlinear phenomena in 
   deep sea waves \c{nlsApply,hasegawa,rog1D}. However, unfortunately since this equation
is defined in $(1+1) $ space-time dimensions,
 the  nonlinear
evolution along  a line only  is possible  to describe by this model.
  Therefore there remains  a persistent 
 need for the  extension  of the NLS equation to
higher dimensions preserving its   integrable structures,
for modelling  more realistic nonlinear phenomena.
An important example of such an  event is  the ocean rogue wave,
  one of the mysteries of nature, which is a
  genuine 2D surface wave appearing in the deep sea.

Unfortunately, until very recently, neither
 a satisfactory $(2+1) $-dimensional
integrable NLS model with   local and regular
 nonlinear interaction of the basic field
 nor an analytic  model for the
 2D ocean rogue wave  
  was  available. 
 A straightforward generalization of the 1D NLS to 2D keeping
its  local cubic nonlinearity   does not serve the required
purpose, since  such a higher dimensional generalization 
 turns out to be a nonintegrable system with no stable soliton solution
\c{2dNLS}.  Zakharov on the other hand has proposed long ago
 an integrable  $(2+1) $-dimensional variant of the  NLS equation 
by   constructing certain form of  Lax operators  by an innovative
trick. However,  unfortunately such an equation is coupled to 
 another potential  field,  expressed  through 
 the  basic fields only in  a  nonlocal way \c{zakhar}.
 Similar  situation  arises also  in the
Davey-Stewartson equation, where again additional  fields are 
coupled  in the  interacting term, which could be 
  related to the basic fields only through 
  nonlocal transformations
   \c{DSE}. Strachan had proposed
another way of constructing higher-dimensional  equations as a
reduction of the self-dual Yang-Mills equation and 
reproduced  the  2D coupled  NLS   equation of Zakharov
and its integrable hierarchies \c{Strach}.

The nonavailability of a higher-dimensional
 integrable NLS equation seems to
make the situation rather desperate, since  the deep-sea  rogue
wave,  clearly a 2D surface ocean wave,
 is described popularly
 by the Peregrine breather   of the 1D NLS equation \c{Peregrine,rog1D},
 which is   a rational  
solution describing wave formation  along an one  dimensional line only. 
In the background of this situation, we have    proposed  recently  a novel
  $(2+1) $-dimensional NLS equation  
 with a local
nonlinear interaction in the basic field. The integrability of this equation   is
achieved when the traditional cubic  amplitude-type  nonlinearity  is replaced by  
a current-like nonlinear term \c{2DNLSarxiv12},
 producing a realistic and exact 2D model for
the ocean   rogue wave \c{rog12}.
However, many  important   properties  
 of this recent $(2+1) $-dimensional integrable
system and its related structures remain unexplored.
Therefore, 	the investigation of this $(2+1) $-dimensional 
NLS equation  regarding the aspects    
 like,  insight into its associated Lax operators, 
related infinite set of  conserved quantities, 
 integrable hierarchy,  explicit construction of
higher soliton solutions and its possible connection with other integrable
  models etc., is our main goal here.

The arrangement of the paper is as follows. In sec. 2 we briefly list the
basic facts about the 1D NLS equation.  Sec. 3 presents the  integrable
2D NLS model together with the related higher conserved quantities
  and investigates  the associated  Lax operators,
 exploring  its relation  with other known models. Sec. 4
derives the soliton  and its higher forms through Hirota's bilinearization.
 The next section
followed by concluding remarks and bibliography, gives some account of the
potential  physical applications of this model.    
\section{Brief account on     integrable  1D NLS equation}
 The well known  integrable 1D
NLS equation expressed through a complex field $q(x,t) $: 
\begin{equation}
 iq_{t}=q_{xx}+2|q|^2q, \label{1dnls}  \end{equation}
together with its complex conjugate equation,
 occupies   an important position among the integrable models due to 
its natural appearance in various applicable fields and its 
rich integrable structure. This nonlinear integrable system is associated
with a matrix form of the linear  Lax equations
\c{solit}:
 \begin{equation}
 \Phi_{x}=U_1(\lambda) \Phi, \ \Phi_{t}=V_2(\lambda) \Phi 
 , \label{LaxEq}  \end{equation}
with the Lax pair
 \begin{equation} U_{1}(\lambda)= i[\lambda \sigma_{z} +
q \sigma_{+} +q^*\sigma_{-}] \label{U1}\end{equation}
and  \begin{equation}
V_{2}(\lambda)=i[V^{(2)}_{11}\sigma_{z}+ V^{(2)}_{12}\sigma_{+} +
 V^{(2)}_{21}\sigma_{-}]
 \label{V2}\end{equation}
where
$V^{(2)}_{11}= 2\lambda^2 - |q|^2  ,
 V^{(2)}_{12} =(V^{(2)}_{21})^*= 2\lambda q -iq_{x},
 $.
\\
considering $\la $ to be real and  
 $ \sigma_{\pm}, \sigma_{z}
$ are Pauli matrices 
$$                  \sigma_{+}=
\left(
\begin{array}{cc}
0 & 1  \\
0 & 0 \\
 \end{array}
\right),  \  \sigma_{-}=
\left(
\begin{array}{cc}
0 & 0  \\
1 & 0 \\
 \end{array}
\right),  \  \sigma_{z}=
\left(
\begin{array}{cc}
1 & 0  \\
0 & -1 \\
 \end{array} \right).$$ 
The subscripts in $x $ and $t$ denote partial
derivatives in the space and time variables, respectively, with the space Lax
equation
 in (\ref{LaxEq}) representing  a scattering problem with $\lambda$ as the
spectral parameter and the field $q(x,t) $ as the scattering potential.  The
subscripts is the Lax pair
 $U_{1}, V_{2}$ on the other hand  designate the maximum power of the
 spectral parameter $\lambda $ contained in it, as may
be checked from their explicit form (\ref{U1}, \ref{V2}). The compatibility 
of the Lax equations (\ref{LaxEq}): $\Phi_{tx}=\Phi_{xt} $ leads to the 
flatness condition of the Lax pair:  
\begin{equation}
U_{1t} - V_{2x} + [U_1,V_2] = 0
,\label{flat}\end{equation}
which yields the 1D NLS equation (\ref{1dnls}) at the $\lambda^0 $ power,
while other  relations at $\lambda^n, \ n=1,2,3  $ are
trivially satisfied. The  association of 
 the 1D NLS equation with a  Lax pair
and its  equivalence   to the
flatness condition of the Lax operators may be 
 considered  as the criteria   of its integrability.
 In fact the space Lax equation $\Phi_{x}=U_1(\lambda) \Phi $
can be used to find analytic solutions of the 1D NLS equation
 through inverse scattering method, yielding also the N-soliton solution
as reflection-less potentials \c{solit}.
 In addition the same equation can be used through the Riccati equation 
 for constructing 
 the infinite set of its conserved quantities $c_j, \
j=1,2,3,\ldots  $ given in the explicit form for 
$x \in [-\infty, +\infty ] $   as
 \bea 
&& c_1= \int dx |q|^2
, \  c_2= i\int dy (q^*_xq-q^*q_x) \nonumber \\
&& c_3 = \int dx (q^*_xq_x+|q|^4) \ll{cn}\eea 
and so on. If we consider $c_3 \equiv H $ as the Hamiltonian 
of the system, one can derive 
  the NLS equation (\ref{1dnls}), which on the other hand 
can also be obtained from the flatness condition (\ref{flat})
 of the  Lax pair 
$(U_1,V_2) $, as mentioned above.      
 If on the other hand we take the next higher 
conserved quantity  $c_4 \equiv H $ as the Hamiltonian 
we get  the higher order
NLS equation
\begin{equation}
 q_{t} + q_{xxx} - 6|q|^2q_{x} =0. \ll{h1nls} \end{equation}
Interestingly, the same nonlinear equation can be obtained as the flatness
condition of the Lax pair $(U_1 (\la ),V_3 (\la )) $ where $U_1(\la) $
is  the same space-Lax operator (\re{U1}), while the time-Lax
operator $V_3 (\la ) $ may be  expressed in an  explicit form  as 
   \be V_{3}=i[V^{(3)}_{11}\sigma_{z}+ V^{(3)}_{12}\sigma_{+} +
V^{(3)}_{21}\sigma_{-}] \ll{V3}\end{equation} Where
 \bea V^{(3)}_{11}=
 [-4\lambda^3 +2 \lambda |q|^2 +i(qq_{x}^*-q^*q_{x})],
\nonumber \\
 V^{(3)}_{12} = (V^{(3)}_{21})^*
= [-4\lambda^2 q + 2i\lambda q_{x}-2|q|^2q + q_{xx}].
 \ll{Vij}\eea
Similarly the infinite set of conserved quantities $c_j, \ j=5,6,\ldots $
 can generate 
infinite number of integrable higher order  1D NLS equations
 forming the 
integrable NLS hierarchy.
 These equations can be shown to be 
  equivalent  to the
flatness condition of the Lax pair  $(U_1 (\la ),V_j (\la )) $ where $U_1(\la) $
is  the same space-Lax operator (\re{U1}),
while   higher   time-Lax
operators $V_j (\la ) $ are  the $j $-th order polynomials in 
  parameter $\lambda $ with matrix coefficients containing 
  higher order derivatives and nonlinearity in the basic field
(with the scaling dimension of each term being  $j $ \c{nonlin}).
  Note however that in spite of the involved 
structure of the higher NLS equations
 their soliton solutions have very similar
form with the simplest 1-soliton  for all the equations given in the form 
\begin{equation}
 q={\rm sech}\eta(x-vt)e^{i(kx+\omega t)} \ll{1DnlsSol} \end{equation}
with only the constant soliton velocity $v $ and the modulation frequency $
\omega $  changing for the hierarchal equations in a particular way.

Among many applications of the soliton solutions of the 1D NLS equations 
the solitonic communication in fiber optics, localized structures in
nonlinear plasma, deep  water waves are worth
mentioning. A rational   solution of the 1D NLS,
 known as the  Peregrine breather
is often used for modelling the rogue waves \c{rog1D}. However Ocean rogue wave
being a 2D surface wave, modelling it using a 1D Peregrine breather 
is indeed unsatisfactory.
  Consequently, there is  a strong need for  a 2D analytic rogue wave 
solution having tunable parameters for amplitude, steepness, velocity etc.
We  present below 
 the  2D NLS  equation  recently proposed by us \c{2DNLSarxiv12,rog12}, 
with  a 2D  solution as model 
for the ocean rogue wave, where the above  realistic requirements  are mostly taken care
of.  
\section{Integrable  2D NLS equation}

The need for constructing a 2D generalization of the NLS equation 
boosted many attempts in this direction. 
  In some studies a straightforward 2D
extension of the NLS equation  is used in the form \c{2dNLS} \begin{equation}
 iu_{t}=u_{xx}-du_{yy}+2|u|^2u \ll{NLS2D} \end{equation} Where $d$
is a  constant.   This
equation however turns out to be a nonintegrable system,
with unstable soliton solutions and   with   
solutions  obtained only  numerically. 

Davey-Stewartson equation (DSE) however is  a successful attempt for constructing
a genuine $(2+1) $-dimensional integrable system, one of its several forms
may be   given by 
the coupled equations
\bea
    i q_t + c_0 q_{xx} + q_{yy} = c_1 |q|^2 q + c_2 q \phi_x,\,
\nonumber \\
    \phi_{xx} + c_3 \phi_{yy} = ( |q|^2 )_x.\,
\ll{DSEq}\eea
with a 
 Lax representation  \c{DSE}. We may note however that using the first  DSE
and its conjugate we can express the external potential $\phi_x $ through
the basic field $q $ and its derivatives, in a complicated but local way. 
This could be achieved for example, by multiplying the first equation by $
q^*$ and its conjugate by $ q$ and then adding them. 
Thus we could remove the external field from both the equations of the DSE,
giving an equation for $q$ in $(2+1) $ dimensions together with
 an additional constraint equation.  This feature of 
  higher dimensional integrable  equations to have an additional
differential constraint seems to be rather universal, as supported by the
available examples and also appears in constructing our 2D integrable NLS
equation. 
In some cases 
additional potential fields  can  be  expressed
more simply through  the basic fields, though with
  nonlocal interactions and with 
  dromion fields sitting at the space
boundaries \c{DSE}.

Another interesting 2D generalization of the NLS equation was proposed by 
Zakharov \c{zakhar}:
\begin{equation}
i q_{t} - q_{xy} +V q =0, \ V_x=2 (| q|^2)_y \ll{zakharEq} \end{equation}
 
derived as a zero curvature condition from an innovatively constructed 
Lax pair \c{zakhar}.
As we see from (\ref{zakharEq}) this coupled NLS type equation 
involves again an additional potential field, which could be 
 related non -locally to the basic 
NLS field. At the same time the potential field 
 $V$ can also be  expressed in local terms
through the basic field and thus can be  removed from both the equations using  
 the conjugate field,  similar to  the DSE as explained above. This would result to an
equation together with an differential constraint involving all  variables 
$x,y,t $, without any external potential. We will see
similar situation for our proposed $(2+1) $-dimensional NLS equation.

Through the reduction of self-dual Yang-Mills equation 
Strachan has introduced a general scheme for constructing higher 
space-dimensional models and re-derived the Zakharov equation 
(\ref{zakharEq})  giving a geometrical meaning to it, together with its
integrable hierarchies,
involving more and more constraint equations.
.   
 
 Recently,    a  $(2+1) $-dimensional integrable  NLS equation
was  introduced,  with 
 local nonlinear interaction and without any external potential field
\c{2DNLSarxiv12}. In this 2D NLS equation 
  the traditional  cubic {\it amplitude}-type nonlinearity in (\re{NLS2D}) is
replaced by a {\it current}-like nonlinearity:
 \begin{equation}
 iq_{t}=q_{xx}-dq_{yy}+2iq(j_{x}-\sqrt{d}j_{y})
\ll{NLS2Di}\end{equation}
with the current terms  defined as 
$\  j_{a}=qq_{a}^*-q^*q_{a}, \ a\equiv x,y $.
 Interestingly, the replacement of such local
 nonlinearity turns the
 nonintegrable equation  (\re{NLS2D}) to an  integrable system
 (\re{NLS2Di})
with all its characteristic properties like the Lax pair, infinite set of
conserved quantities, higher soliton solutions etc., which we will 
present below  sequentially.

Before proceeding further we rewrite equation (\re{NLS2Di}) in a more
compact form by rotating the coordinate  frame  on the
plane by an angle $\frac \pi 4$ together with a 
 scale transformation:
 $ (x,y) \to (\bar x,\ \bar y) $ with
$ \bar x=\frac 1 2(-x+\frac 1 {\sqrt{d}}y),\ 
\bar y=\frac 1 2(x+ \frac 1 {\sqrt{d}} y)  $ and $ \bar t= 2t$ and a 
 scaling
of the field,   
yielding 
\be iq_{t}+q_{xy}+2iq(qq_{x}^*-q^*q_{x})=0 \ll{NLS2Do}
\end{equation}
where  the $bar $  over the coordinates is   omitted for 
 the sake of notational simplicity.

\subsection{ Lax operators for the 2D NLS equation}
We introduce here the Lax pair associated with the 
the  integrable  2D NLS equation (\re{NLS2Do}). 
Since the idea is to generalize the NLS equation 
  to $x,y,t $ coordinate variables, preserving the integrable
structure of the system, we  
 start by generalizing  the pair of Lax  equations (\re{LaxEq})
to   
 \begin{equation}
  \Phi_{y}=U_2(\lambda) \Phi, \ 
\Phi_{t}=V_3(\lambda) \Phi 
 , \label{LaxEq3}  \end{equation}
where  we will 
use the  pair $U_2(\lambda),V_3(\lambda)  $ for constructing the 
2D generalization of the NLS, by
   taking them  in the same form 
as in (\re{V2},\re{V3}),  considering $U_2(\lambda)\equiv 
V_2(\lambda). $    It is interesting to observed that the Lax operators
   taken usually 
 in a rather complicated element-wise
 form as presented above, 
can be expressed in more compact and interrelated matrix form
as
\bea  
  U_2(\la)=2\la U_1(\la)+U_2^{(0)}, \ \  U_1(\la)=i(\la \sigma^3+U^{(0)}), 
 V_3(\la)=2\la U_2(\la)+V_3^{(0)}. \ll{V23}\eea
  We  notice 
 that  the higher order Lax
operators,  
 finding of   which is known to be  
 a difficult task in general, could  be simplified by  expressing them 
through those of the lower orders, except the
$\lambda$-independent terms $U^{(0)}, \ U_2^{(0)}, \ V_3^{(0)} $. 
We choose these crucial terms  as
\bea
U_2^{(0)}=\sigma^3(U^{(0)}_x-i{U^{(0)}}^2), \
V_3^{(0)}=D(U^{(0)})-[U^{(0)},U^{(0)}_x], \ 
\mbox{with } \ \ U^{(0)}=
\left(
\begin{array}{cc}
0 & q  \\
 q^* & 0 \\
 \end{array}
\right),
\ll{Lax}\eea 
where for 
the standard  choice: 
\be  D(U^{(0)})= iU^{(0)}_{xx}+2i{U^{(0)}}^3 \ll{v30}\ee
we can recover the Lax operator forms including (\ref{V3}, \ref{Vij}),
well known for the NLS hierarchy.
 
We notice however that The $\lambda $-independent parts of the 
Lax operators  should be  obtained 
 additionally and could be  constructed guided   by the 
 principle of maintaining  
the  scaling dimensions \c{nonlin}, which therefore 
seems to give   some  flexibility in their  choice. We
will exploit this   freedom for constructing a 
different  Lax operator. In
particular by choosing   the matrix  $ D(U^{(0)}) $
  in  $V_3^{(0)} $ differently,   we can generate 
different  constraint equations.

If we start from the linear system (\ref{LaxEq3}), 
considering the 
Lax operators (\ref{Lax}) with the  standard choice (\ref{v30}), 
the compatibility of the system leading to  the flatness condition on the 
 Lax pair:  $(U_{2}, V_{3})$  
would produce  different equations 
at different powers of the spectral parameter $\lambda . $  
At  $\lambda^2$  we get
a  NLS like equation, 
involving only variables $x,y $:
 \be iq_{y}=q_{xx}+2|q|^2q, \label{1dnlsy}  \end{equation}
which being  not an evolution equation
 may be considered as a nonholonomic differential 
 constraint on the field.
At   $\lambda$ we obtain the integrable 2D NLS equation 
(\re{NLS2Do})
together with its complex conjugate and finally   
 at
$\lambda^0$  we get yet  another  nonlinear    equation 
\be   q_{xt}+q_{ y  y}  +2i|q|^2q_y+2q_x(qq^*_{x}-q^*q_{x})=0,
 \ll{2dsInlsE} \ee
along  with 
\be
i(|q|^2)_{t}+(q^*q_{xy}-q^*_{xy}q)=0 .
 \ll{NLSm2Do}\ee Note that we may consider the 2D NLS equation 
(\re{NLS2Do}) as our main
equation, together with the differential constraints  (\re{1dnlsy}) and
(\re{2dsInlsE}), since the additional equation (\re{NLSm2Do}) 
  is  derivable from  equation  (\re{NLS2Do}) and its
conjugate and hence  not an independent equation.
We observe therefore that though we can derive 
the integrable  2D NLS equation (\ref{NLS2Do}) from 
the Lax pair 
$(U_{2}, V_{3})$ it give also  two constraint equations 
(\ref{2dsInlsE}) and (\ref{1dnlsy}).
However since 
too many constraints for a single equation is an undesirable situation,
we wish to seek now a way for the reduction of such constraints. 

 Intriguing, if we  choose  
the matrix  $D(U^{(0)}$ occurring in  (\ref{Lax}) different from
 the standard  one (\ref{v30}) by taking  simply  
\be  D(U^{(0)})= -U^{(0)}_{y}, \ll{v300}\ee
preserving the   scaling dimension, the constraint 
(\re{1dnlsy}) disappears  miraculously as we wanted 
and the Lax pair $U_2,V_3 $ 
 with the important  modification   (\ref{v300}) would  yield 
 the 2D NLS equation
 (\re{NLS2Do})
:
$$ iq_{t}+q_{xy}+2iq(qq_{x}^*-q^*q_{x})=0 
$$
together 
with  a single  higher order constraint equation (\re{2dsInlsE}).
Note that using further the evolution equation (\re{NLS2Do})   
 this constraint  can be reduced  to a nonevolutionary
type  constraint equation involving only derivatives in $x $ and $y$:
 \be iq_{xxy} +q_{ y  y}  
+2i|q|^2q_y-2q(qq_{xx}^*-q^*q_{xx})=0
 \ll{const}\ee
 Therefore, we may conclude, that the 2-dimensional
integrable NLS  equation (\re{NLS2Do})
 with local nonlinear interaction and 
without introducing  any external potentials is possible to construct 
from an associated  Lax pair as explained above. However this equation
has  a single differential 
constraint (\ref{const}), common  also 
for   the   Zakharov equation  and  the  DSE. 
We will be concerned here mainly with the integrable 2D NLS equation 
(\re{NLS2Do}) treating it as an independent equation,
since unlike   the DSE 
(\ref{DSEq}), the Zakharov  equation (\ref{zakharEq})
or the Strachan's construction
 \c{Strach},  it does not have any potential field
linked to the constraints.

\subsection {Relation with Zakharov and Strachan construction
 } Using the twistor space technique Strachan has generated a series of
$(2+1) $-dimensional integrable models belonging to the NLS  family, e.g. 
Zakharov equation having  an external  potential with one constraint, its higher order flow
involving two additional potentials and two constraints etc.  as well as a similar 2-dimensional generalization
of the derivative NLS equation involving also external potential fields \c{Strach}. 
In the scheme proposed by Strachan for generating $(2+1) $-dimensional integrable
equations, the set of linear systems for some particular choice of parameters
is given
   by 
\bea \Phi_{x_1}= (-\lambda A_0+D_1)\Phi, \ 
\Phi_{x_2}= \lambda \Phi_{x_1}+D_2\Phi, \nonumber \\ 
\Phi_{x_3}= \lambda \Phi_{x_2}+D_3\Phi, \ 
\Phi_{x_4}= \lambda \Phi_{x_3}+D_4\Phi, \ll{strach}\eea type 
etc.  For establishing connection  of the linear system (\ref{strach}) with our 
Lax equation (\re{LaxEq3}) given through 
 the Lax operators  (\re{V23},\re{Lax}),
 let us  take the first  three equations in (\re{strach}) and 
 denote $x_1= x $, \ $x_2= y $, \ $x_3= t $. 
Denoting further
$A_0=-i \sigma^3, \  D_1=iU^{(0)} $
one gets the standard Lax equation $ \Phi_{x}=U_1(\lambda) \Phi $ from 
the first Strachan relation in (\re{strach}) where $U_1 $
is given as in (\re{V23}). Using this Lax equation 
and denoting $ D_2=U_2^{(0)} $  we 
 recover the  Lax equation $ \Phi_{y}=U_2(\lambda) \Phi $  
 from  the
second linear system of (\re{strach}), with $ U_2(\la)=2\la U_1(\la)+U_2^{(0)}, $ \ as 
 in
(\re{V23}). In a similar way by replacing  $ D_3=V_3^{(0)} $ and using 
the second Lax equation together with the form  $ V_3(\la)=2\la U_2(\la)+V_3^{(0)}$
 as in (\re{V23}),  one obtains from the third equation in the Strachan
series  (\re{strach})  our time-Lax equation 
$ \Phi_{t}=V_3(\lambda) \Phi $. This shows an intimate relation of our model
 with the Strachan construction revealing  a  geometrical meaning 
through twistor space formalism and self-duality construction of the
 $(2+1) $-dimensional integrable NLS equation (\ref {NLS2Do}) we are
investigating here. However a note of caution is that the explicit form for
our matrices $V_2^{(0)}, V_3^{(0)} $ containing only the basic fields are
different from the Strachan construction $D_2,D_3 $ involving external
potentials.

In the family of   integrable equations in $(2+1) $-dimensions together with their
 hierarchies constructed by Strachan including  the Zakharov equation (\ref {zakharEq})
there  involves  an
external potential field  $V(x,y,t), $
 expressed nonlocally through the basic
NLS field as
 $ V(x,y,t)=2 \int^x_{-\infty} dx (| q|^2)_y +f(y,t), $
with arbitrary function $f(y,t) $ defined at  $x \to -\infty$-boundary. 
We  can find therefore a link 
of the Zakharov equation to the 
2D NLS   (\re{NLS2Do}),  if we   additionally consider the NLS-like
constraint   (\re{1dnlsy}) together with  
 its conjugate deriving the relation $ 
 (| q|^2)_y=i(qq_{x}^*-q^*q_{x})_x  .$ Comparing with the Zakharov equation
one gets a particular solution for the potential field as $V=
i(qq_{x}^*-q^*q_{x})$ for the choice $f(y,t)=0 $, which from the first
of the Zakharov's equations yields the 2D NLS equation
(\re{NLS2Do}). 

 \subsection{ Infinite set of conserved quantities}
 Systems with infinite degrees of freedom like the 2D-NLS equation
(\ref{NLS2Do}), when integrable, 
should have infinite set of independent  conserved quantities.  
We  generate here the related infinite set of  conserved charges  $C_n,
 \ n=1,2, \ldots $ in the
explicit form, demonstrating  again an important
feature of the  2D NLS equation  
linked to its integrability. 
In analogy with the 1D NLS equation we start  from the  
linear system (\re {LaxEq3}), but 
 use now the Lax equation along the $y $-direction:
   $\Phi_y=U_2(\lambda)\Phi. $ 
Note, that 
for the wave function 
 $\Phi(\lambda,y)=(\phi, \tilde \phi), $ the component 
 $$ \phi(y,\lambda)=e^{\int_{-\infty}^y \rho(\lambda,y') dy'},$$ with $ \ 
\int^{+\infty}_{-\infty} dy' \rho(\lambda,y')= \sum_{n=1}^\infty C_n
 \lambda^{-n}$
 acts as a generator of the conserved quantities, yielding 
 $$ {\rm ln}
\phi(y=\infty, \lambda)=\sum_{n=1}^\infty  C_n \lambda^{-n}.  $$
Therefore using $U_2(\lambda)$ as in (\re{V23}) or in more 
 explicit  form (\re{V2}) in  the Lax equation  
$\Phi_y=U_2(\lambda)\Phi, $   
we can build systematically the 
 infinite set of conserved charges: $C_n, \ n=1,2, \ldots $\ \
through a recurrence relation giving 
 \bea C_1= i\int dy (q^*q_x-q^*_xq),   \  C_2=\int dy
  (i\frac 1 2 (q^*_{y}q -q^*q_{y})+ q^*_{x}q_{x} + | q|^4
) , \ C_3=\int dy (q^*_{y}q_x+q^*_{x}q_y) ,\nonumber \\  
   C_4= \int dy \left(
iq^*_{xy}q_{x}+q^*_{y}q_y-
i    |q|^2(q^*q_{y}- q^*_yq)
 -2 |q|^2 \ q^*_xq_{x}+( {q^*}^2q^2_{x}+
{q_x^*}^2q^2
 )\right),\ll{2dCn}\eea
and so on.
We note the involvement of both the space-variables $x,y $ in this series of 
independent conserved quantities, which also gives another 
strong argument in favor of the integrability of the 2D nonlinear 
 equation (\re{NLS2Do}). Taking these conserved quantities as Hamiltonians
$H\equiv C_n $ we can generate the integrable hierarchy for this   
 2D NLS equation.

\section{Soliton  solutions} Recall that the integrable nonlinear equations
allowing linear spectral problem can be solved for the general initial value
problem by the inverse scattering method, to obtain in particular the exact
soliton solutions \c{solit}.  However this method has been developed mostly
for systems like the NLS, KdV, mKdv, sine-Gordon equations etc., belonging
to the AKNS spectral problem  or like the derivative NLS equation.
  For higher dimensional equations
like DS equation the associated Lax operator on the other hand  usually
contains no spectral parameter and needs therefore a different treatment
 for their solution \c{DSE}.  Noticing that the associated spectral problem given
through the Lax operator $U_2(\la) $ (\re{V2}) for our 2D NLS equation, does
not fall into any of these well known problems,  we resort to a
direct method through Hirota's bilinearization for extracting exact
solutions to the $(2+1) $-dimensional nonlinear equation (\re{NLS2Do}). 
 As it is known, successful application of the Hirota's bilinearization
method should yield an exact 1-soliton and subsequently, 
through a recursive method the higher soliton
solutions can also be obtained.
 For expressing the  2D NLS equation in the   bilinear form we use the
standard  transformation
\begin{equation}
 q(x,y,t)= \frac{G(x,y,t)}{F(x,y,t)}, \ll{GF} \end{equation} where
$G(x,y,t)$ and
$F(x,y,t)$ are complex and real functions, respectively.  Inserting (\re{GF})  
in (\re{NLS2Do}) one derives   the  pair of bilinear
equations:
 \begin{equation}
 i(FG_{t}-GF_{t}) + (FG_{xy}+ GF_{xy}-G_{x}F_{y}-G_{y}F_{x})=0, \ll{Hirota1}
\end{equation}
\begin{equation}
 2i(GG_{x}^*-G^*G_{x}) + 2(F_{x}F_{y}-FF_{xy})=0.
\ll{Hirota2}\end{equation}
Following  the standard prescription of a formal  series expansion:
\bea
 F = 1 + \epsilon^2 F_{2} + \epsilon^4 F_{4} + \cdots \ll{F123}\\
 G =  \epsilon G_{1} + \epsilon^3 G_{3} +  \cdots,  \ll{G123} \eea
where   $\epsilon$  need not  be small,
  we obtain the following equations
corresponding to different orders in $\epsilon$.
\begin{equation}
{\rm O}(\epsilon):\ \  iG_{1t} + G_{1xy}=0
\ll{G1}\end{equation}
\begin{equation}
{\rm O} (\epsilon^2): \ \  2F_{2xy}=2i[G_{1}G_{1x}^* - G_{1}^*G_{1x}]
\ll{F2}\end{equation}
\begin{equation}
{\rm O} (\epsilon^3): \ \ iG_{3t}+G_{3xy}]=i[G_{1}F_{2t}-F_{2}G_{1t}]-[F_{2}G_{1xy}+G_{1}F_{2xy}-G_{1x}F_{2y}-G_{1y}F_{2x}]=0
\ll{G3}\end{equation} \begin{equation}
{\rm O} (\epsilon^4): \ \  2F_{4xy}=2i[G_{3}G_{1x}^*
+G_{1}G_{3x}^*-G_{3}^*G_{1x}-G_{1}^*G_{3x}]+2[F_{2x}F_{2y}-F_{2}F_{2xy}]
\ll{F4}\end{equation} and similarly  higher order equations.

  \subsection { 1-soliton}
To construct    1-soliton solution for 
(\re{NLS2Do}) we assume the ansatz
\begin{equation}
 G_{1}=e^{\eta_{1}}, \ 
 \eta_{1}=k_{1}x + p_{1}y-w_{1}t+\eta_{1}^{0}\ll{G11}
\end{equation}
where $k_{1}, p_{1},w_{1},\eta_{1}^{0}$ are complex constants.
 From equation (\re{G1}) therefore one obtains
 the associated dispersion relation
$\  w_{1}=-ik_{1}p_{1}, \ $
using which  the equation  (\re{F2}) is solved easily  to yield 
\begin{equation}
 F_{2}=i(k_{1}^*-k_{1})
\frac{e^{(\eta_{1}+\eta_{1}^*)}}{(p_{1}+p_{1}^*)(k_{1}+k_{1}^*)}. \ll{F22}
\end{equation} 
We can verify using  (\re{G11}) and  (\re{F22}), that
 all  higher order terms in $\epsilon$ like (\re{G3}, \re{F4}) etc.,
beyond $G_{1}$ and $F_{2}$  trivially vanish.
  Absorbing $\epsilon$ in   arbitrary
 constant  $\eta_{1}^{0}$, 
we construct from  
  (\re{GF}) using  (\re{G11}) and  (\re{F22})
the 1 soliton solution in the form 
\begin{equation}
 q(x,y,t)=\frac{G_{1}}{1+F_{2}}=\frac {e^{\eta_{1}}}
 {1+ \alpha 
e^{(\eta_{1}+\eta_{1}^*)}}
\ll{1sol0} \end{equation} 
where $\alpha$ depends on the parameter $k_{1}, p_{1} .$ 
If additionally we use the constraint  equation (\re{const}), 
through its   dispersion relation together with that of the main equation   
we can link 
the independent parameters as  $ p_{1} = -i k_{1}^2$, simplifying the 
soliton solution (\re{1sol0}) to
 yield the conventional form \be q(x,y,t)
={\rm sech}\xi \ e^{i\theta}, \ \mbox{with} \ \xi=\eta(x+v_y y+v t), \
\theta=(k_xx+k_yy+\omega t).  \ll{1sol2}\ee where all the soliton
parameters $\eta, v_y, v, k_x, k_y, \omega $ can be expressed through two
independent real spectral parameters $\lambda=k_x+i\eta $.  Note, that the
wave front of the 2D soliton along a line travels in time as an exact
solution of the $(2+1) $-dimensional NLS equation (\re{NLS2Do}).  Such
solitons are often called a line-soliton.  We unfortunately could not find a
dromion-like exponentially decaying moving lump-soliton as found in DSE
\c{DSE}.  Note that the Zakharov equation also exhibit
 no dromion solution as reported in \c{radha}.       
A frozen picture of the modulus of our travelling  soliton solution (\re{1sol2}) at
time $t=0 $  is shown in Fig. 1.

\begin{figure}
\includegraphics[height=5cm,width=7cm]{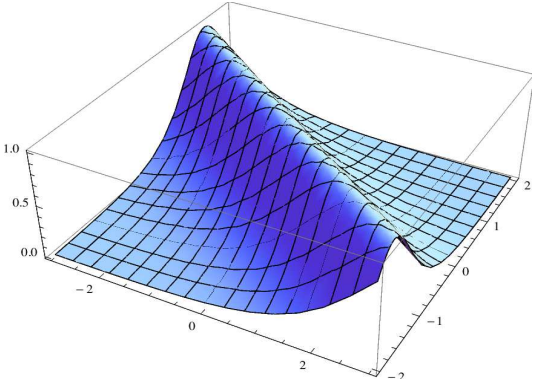}
\caption{Modulus of 1 soliton with $k_{1r}=1,k_{1i}=1,\eta_{1r}^{0}=1,
\eta_{1i}^{0}=-1 $ and at t=0} 
\end{figure}
\subsection{2-Soliton}

For  obtaining  2-soliton solution we start with  the standard procedure
assuming  
\bea
 G_{1}=e^{\eta_{1}}+ e^{\eta_{2}},
\ \ \mbox{with} \nonumber \\
 \eta_{1}=k_{1}x + p_{1}y-w_{1}t+\eta_{1}^{0}
, \  \eta_{2}=k_{2}x + p_{2}y-w_{2}t+\eta_{2}^{0} 
,
\ll{G12}\eea
 where the parameters   involved are complex numbers. 
 Applying similar   dispersion relations as earlier 
we get 
$ w_{1}=-ik_{1}p_{1}, \ 
 w_{2}=-ik_{2}p_{2} $ and obtain from (\re{F2})
 \begin{equation}
 {F_{2}=[e^{(\eta_{1}+\eta_{1}^*+R_{1})}+
 e^{(\eta_{2}+\eta_{2}^*+R_{2})}}
+e^{(\eta_{2}+\eta_{1}^*+\delta_{0})}+
{e^{(\eta_{1}+\eta_{2}^*+\delta_{0}^*)}]},
\ll{2F2}\end{equation}
where all the constant parameters can be worked out explicitly
 (see Appendix I
).
Similarly
 equation (\re{G3}) at  higher order expansion   gives 
\be {G_{3} = e^{(\eta_{1}+\eta_{1}^*+\eta_{2}+ \delta_{1})} 
+ e^{(\eta_{1}+\eta_{2}^*+\eta_{2}+ \delta_{2})}},
\ll{2G3}\end{equation}
 where the relevant 
 parameter details 
are given in  Appendix II.  
 Using  further equation (\re{F4}) one obtains
\begin{equation}
  F_{4} = e^{(\eta_{1}+\eta_{1}^*+\eta_{2}+\eta_{2}^*+R_{3})},
\end{equation}
 with the relevant parameters
 presented in  Appendix III. Note that in all the expressions of $F_2,G_3 $
and $F_4 $  
the   two-soliton interaction is explicit. 
For simplifying the expressions,as mentioned earlier, 
we can use the constraint equation (\re{const}),
imposing the relations between
 $k_{1}$ , $p_{1}$ and $k_{2}$ , $p_{2}$ as $p_{1}=-ik_{1}^2$ ,
 $p_{2}=-ik_{2}^2$ (see Appendix IV).  Here we find again, that the higher
 order terms in $\epsilon $ beyond $G_{3}$ and
$F_{4}$ trivially vanish, leaving the exact 2- soliton solution in the form
\begin{equation}
 q(x,y,t)=\frac{G_{1}+G_{3} }{1+F_{2}+ F_{4}}\ll{2sol} \end{equation} A
graphical plot of the  modulus of this solution   in $(2+1)
$-dimensions, frozen at time $t=2$, is shown in Fig.  2, where 
the 2-soliton as two interacting 1-solitons 
 is clearly seen on a 2D $(x,y) $-plane. 

\begin{figure} \includegraphics[height=5cm,width=7cm]{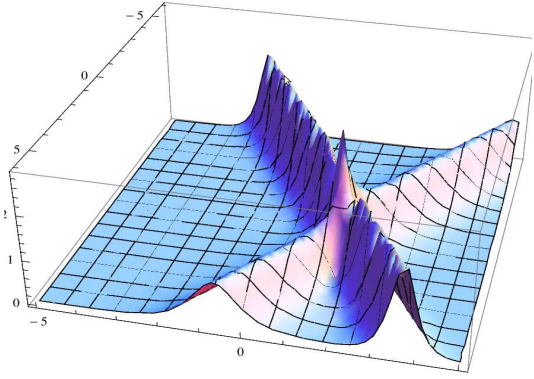}
\caption{Modulus of 2 soliton with
$k_{1r}=1,k_{1i}=1,k_{2r}=2,k_{2i}=-1,\eta_{1r}^{0}=1, \eta_{1i}^{0}=1,
\eta_{2r}^{0}=1, \eta_{2i}^{0}=1 $ and at t=2} \end{figure}

 \section{Applicable aspects of integrable 2D NLS model} Similar to the
well known integrable 1D NLS equation, we expect the NLS equation
(\ref{NLS2Do}) in $(2+1) $ dimensions to have also diverse applications in
different fields.  Moreover, since the 2D NLS model can describe more
realistic nonlinear phenomena in two-dimensional plane, its importance for
physical application should be more significant.  A key reason for the
applicable success of the 1D NLS model is its link to and derivability from
 the basic physical  equations like 
the Maxwell equation in electrodynamics and the Euler equation in
hydrodynamics. It is therefore encouraging that the 2D NLS equation 
(\ref{NLS2Do}) has also found to be derivable under a certain  space-asymmetry
in 2D  from the  Euler equation
\c{rog12} as well as from the Maxwell equation with  physical
conditions  relevant to the plasma physics \c{plasma13}.  
We briefly report below few important application of the model 
 found recently by us \c{rog12,light13}.

\subsection{Ocean rogue wave model based on 2D NLS equation}
Ocean rogue waves  are
 extremely high and steep surface waves, 
appearing suddenly and disappearing  equally fast
 in a relatively calm sea. However due to the lack of satisfactory 2D models
the real phenomena  as well as experimental observations in various fields 
are usually attempted to be described by  
  one dimensional models based on the well known 1D NLS equation
\c{rog1D}
and most popularly by its 1D Peregrine breather solution
\bea q_P(x,t)= e^{-2it}(u+iv), \ u=G-1, \ v=-4tG,
 \nonumber \\
\   G= 1 /F(x,t), \ F(x,t)=x^2+4t^2+\frac 1 4. 
\ll{Pbr} \eea
 The solution (\re{Pbr}) represents a breather mode ${\rm cos} 2t $  
  with unit intensity at both $t\to \pm \infty $, while    at   $t=0 $
  amplitude of the wave rises suddenly ,
 attaining  its maximum at $x=0  $ as shown in Fig. 3

\begin{figure}[!h]
\centering
{
 \includegraphics[width=4.5cm, angle=-90]{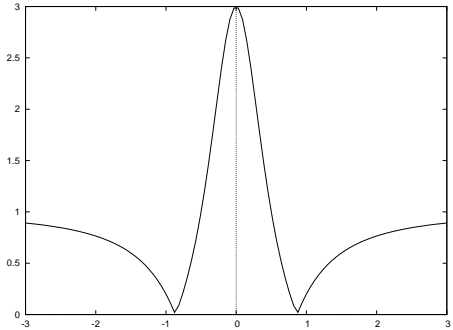}
}

\caption{ Intensity of the full grown 1D rogue wave as modelled by the 
static Peregrine breather $|q_P(x,0)|.$
}
\end{figure}

Notice  however that, 
the  NLS equation (\re {1dnls}) together with its different generalizations   
are   equations only in  $(1+1) $-dimensions
and therefore  all of  their   solutions,  including the  PB and its 
higher order generalizations,  
 can describe  the
 time evolution of a wave only   along an one dimensional  line
(as in Fig. 3). Moreover, 
due to  the absence of any free parameter in (\re{Pbr}) and its higher order
generalizations, 
 the  maximum amplitude and steepness of the  rogue wave  model as well as 
the duration and the   speed of its appearance    
are all fixed. Therefore 
  describing the  actual ocean  rogue waves, parameters of  which may vary continuously 
 from one  event to another, becomes difficult.

Therefore there is an immediate need for a realistic rogue wave model.
We have proposed recently a 2D ocean rogue wave model \c{rog12} 
based on a modification 
of   the integrable 2D NLS equation
with the addition of an  {\it ocean current}   term 
, since the  role of ocean currents in the formation 
of the rogue wave is found to be crucial {\c{prlCurr11}}. This 
modified 2D nonlinear equation
with a specific    form of the ocean current
 is found to yield an interesting 2D dynamical lump solution  \c{rog12}
\bea
q_{P(2d)}(x,y,t)= e^{4ix}(-1 + (1-i4x)\frac 1 {F(x,y,t)}),
\nonumber \\
 F(x,y,t)=  4x^2+\alpha y^2+   \mu  t^2+c, 
\ll{2dPbt} \eea
which can serve as a satisfactory ocean   rogue wave model.
The  solution (\re{2dPbt}) forms into a full grown
lump rogue wave at $t=0$ (see Fig. 4), while disappearing fast  to the background plane wave 
at distant past and  future  ($t\to \pm $).    
Comparison  with the Peregrine rogue wave  (Fig. 3) shows vividly
the  2D nature of our rogue model. The   presence of  
 arbitrary parameters $\alpha,c, \mu $ in    solution   (\re{2dPbt})
plays crucial roles for regulating the maximum amplitude,
steepness, speed and duration of the rogue wave, which is difficult to
achieve
in the conventional   Peregrine type model. 
\begin{figure}
{
 \includegraphics[width=5.cm, angle=-90]{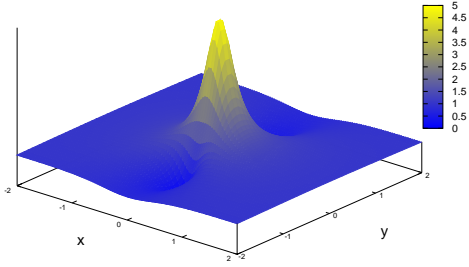}
}
\caption{
Full grown 2D rogue wave  formed at time $t=0.0$ based on the modified 2D NLS
equation
}  
\end{figure}

\subsection{Application to Nonlinear Optics}
 1D NLS equation
is a successful model in nonlinear optics. Apart from  its 
 soliton based optical communication  \c{hasegawa},
 possibility of achieving 
the bending of a light beam based on this model is also proposed recently 
\c{NLtheor}.  
Therefore, a natural expectation is,  that the integrable 2D NLS equation
admitting  stable soliton and other  localized solutions
 should be equally applicable to the
nonlinear optics extending the NLS based models to 2D plane. 
Interestingly, the 2D NLS
equation, like its well known 1D counterpart \c{maimistov}, is found to 
 retain its
integrability when coupled to the self-induced transparency (SIT) equations:
 \begin{eqnarray} &&iE_z+E_{yt}+2iE(E^*E_t-
EE_t^*)=2p, \nonumber \\ && ip_t=2(NE-\omega_0p),\  iN_t=E^*p-p^*E,
\label{2DNLSp} \end{eqnarray}
opening up a novel possibility of more stable 
2D nonlinear optical wave propagation through coherently excited 
resonant medium doped with Erbium  atoms \c{light13}.
Here $ p(t,y,z)$ is the polarization in the
 resonant medium induced by the electric field $ E(t,y,z)$ and $ \ -1 \
\leq N(t,y,z) \leq \ +1 $ is the population inversion
 profile of the dopant atoms. 
  In the set of
equations (\ref{2DNLSp}) the transverse dimension $x$ is
 replaced by  time variable $t$, as customary in nonlinear optics models.
The 
coupled  set of 2DNLS-SIT equations
(\ref{2DNLSp}) turns out  to be
 an integrable system associated with a Lax pair, 
given by a deformation of the 
 NLS Lax operator: $\ V_3(\lambda)
 +\frac 1 \lambda V_{-1},  $ where  the additional matrix
 $ V_{-1} (N,p,p^*)$,
 contains the SIT fields: $ N,p,p^* $.
The set of the coupled integrable equations 
   (\ref{2DNLSp}) 
for the electric field admits soliton solution 
again in the same form
(\re {1sol2}),
though with a different expression for    velocity  $ v $ and 
   modulation frequency $ \omega$. Similarly 
 the other fields $p(t,y,z) $ and
$N(t,y,z). $ also exhibit  soliton solutions. Note that
 the  stable loss free soliton solution of (\ref{2DNLSp}) for the electric
field, valid under the conventional assumption of a constant
 initial condition for the
population inversion: $N(t\to -\infty,y,z)=N_0=-1 , $ would propagate with 
a constant velocity and with an invariant shape and 
 is suitable for information transfer through nonlinear
media and can have applications in 2D processes where transverse direction 
also plays a  prominent role.   
However, intriguing in place of constant initial population
inversion $N_0=-1 $, if we can maintain   
its initial setting at time $t\to -\infty $
  as an arbitrary  function $ \ -1 \
\leq N_0(y,z) \leq \ +1,  $ one   can obtain an accelerating soliton 
 with variable  velocity $v(y,z) $ and   
modular frequency $\omega(y,z) $ \c{light13}.

Existence of such accelerating solitons in an integrable system, apparently
 violating the conventional belief, is explained by the fact that 
due to nontrivial initial condition, the energy is stored initially 
and acts without changing the total energy of the system \c{AIPad11}. 
The application of such novel solitonic features in two-space dimensions
could have innovative applications.
One of such applications is a possibility for bending of the 
 light beams  in 1D and 2D by using integrable  NLS-SIT equations
 have been found recently \c{light13}.   
\section {Concluding Remarks}

The well known NLS
 equation  with wide applications
and rich mathematical structure is an integrable system 
 in $(1+1) $-dimensions. 
A  $(2+1) $-dimensional  integrable extension of this equation proposed by
us recently is investigated here to show  
 similar rich properties for this equation. Apart from analysing 
 its underlying
integrable structure  associated with a  Lax pair and exploring the
infinite set of conserved quantities of this system 
 in the explicit form, we have found
also  an
 exact 1-soliton slution
 along with   its generalizations and shown the relation of this model
 with other known
integrable models in higher dimensions. One of the common properties 
of these higher dimensional models, namely the occurence 
 of nonholonomic
differential constraints, is found also to be  present in our model.

To obtain exponentially localized dromion-like solutions for the present
model, which we have failed to find, as well as 
to extend its practical  applications  at par with the well known NLS 
model, 
 whould be  important future  problems.

\section {Appendix: parametrs related to 2 soliton solution}
\noindent {\bf I. Parameter details for $F_2 $} 
$$e^{R_{1}}=i\frac{(k_{1}^*-k_{1})}{(p_{1}+p_{1}^*)(k_{1}+k_{1}^*)},e^{R_{2}}=i\frac{(k_{2}^*-k_{2})}{(p_{2}+p_{2}^*)(k_{2}+k_{2}^*)},
$$ 
$$ e^{\delta_{0}}=i\frac{(k_{1}^*-k_{2})}{(p_{2}+p_{1}^*)(k_{2}+k_{1}^*)},
e^{\delta_{0}^*}=i\frac{(k_{2}^*-k_{1})}
 {(p_{1}+p_{2}^*)(k_{1}+k_{2}^*)}$$

\noindent {\bf II. Parameter details related to  $G_3 $} 

\begin{eqnarray}
 e^{\delta_{1}}&=&
    \frac{i}{[(k_{2}+k_{1}^*)(p_{1}+p_{1}^*)+(k_{1}+k_{1}^*)(p_{2}+p_{1}^*)]}[\frac{(k_{1}^*-k_{1})
    (p_{2}-p_{1})}{(p_{1}+p_{1}^*)}+\nonumber \\
&+&\frac{(k_{1}^*-k_{1})(k_{2}-k_{1})}{(k_{1}+k_{1}^*)}+
\frac{(k_{1}^*-k_{2})(p_{1}-p_{2})}{(p_{2}+p_{1}^*)}]+
\frac{(k_{1}^*-k_{2})(k_{1}-k_{2})}{(k_{2}+k_{1}^*)}] \end{eqnarray}
\begin{eqnarray}
 e^{\delta_{2}}&=& \frac{i}{[(k_{1}+k_{2}^*)(p_{2}+p_{2}^*)+(k_{2}+k_{2}^*)(p_{1}+p_{2}^*)]}[\frac{(k_{2}^*-k_{2})
    (p_{1}-p_{2})}{(p_{2}+p_{2}^*)}+\nonumber \\
&+&\frac{(k_{2}^*-k_{2})(k_{1}-k_{2})}{(k_{2}+k_{2}^*)}+
\frac{(k_{2}^*-k_{1})(p_{2}-p_{1})}{(p_{1}+p_{2}^*)}]+
\frac{(k_{2}^*-k_{1})(k_{2}-k_{1})}{(k_{1}+k_{2}^*)}]
\end{eqnarray}
\noindent {\bf III. Parameter details for  $F_4 $}
\bea e^{R_{3}}&=&\frac{1}{(k_{1}+k_{1}^*+k_{2}+k_{2}^*)(p_{1}+p_{1}^*+p_{2}+p_{2}^*)}
[\{ie^{\delta_{2}}(k_{1}^*-k_{1}-k_{2}-k_{2}^*)\nonumber
\\
&+&ie^{\delta_{1}}(k_{2}^*-k_{1}-k_{2}-k_{1}^*)+ie^{\delta_{1}^*}(k_{1}^*+k_{2}^*+k_{1}-k_{2})+
ie^{\delta_{2}^*}(k_{1}^*+k_{2}^*+k_{2}-k_{1})\}\nonumber \\
&+&\{(k_{2}+k_{2}^*-k_{1}-k_{1}^*)
(p_{1}+p_{1}^*)+(k_{1}+k_{1}^*-k_{2}-k_{2}^*)
(p_{2}+p_{2}^*)\}e^{R_{1}+R_{2}}\nonumber
\\
&+&\{(k_{2}+k_{1}^*-k_{1}-k_{2}^*)
(p_{1}+p_{2}^*)+(k_{1}+k_{2}^*-k_{2}-k_{1}^*)
(p_{2}+p_{1}^*)\}e^{\delta_{0}+\delta_{0}^*}
]
\end{eqnarray}
\noindent {\bf 
IV. Simplifying expressions for the parametrs $R_i, \ i=1,2,3. $}
 
For simplifying the expressions we can impose  the relations between
 $k_{1}$ , $p_{1}$ and $k_{2}$ , $p_{2}$ as
 $p_{1}=-ik_{1}^2$ , $p_{2}=-ik_{2}^2$, which would yield 
 $$e^{R_{1}}=\frac{1}{(k_{1}+k_{1}^*)^2},e^{R_{2}}=\frac{1}{(k_{2}+k_{2}^*)^2},
e^{\delta_{0}}=\frac{1}{(k_{1}+k_{2}^*)^2},
e^{\delta_{0}^*}=\frac{1}{(k_{2}+k_{1}^*)^2},
$$ $$
e^{\delta_{1}}= \frac{(k_{1}-k_{2})^2}{(k_{1}+k_{1}^*)^2(k_{2}+k_{1}^*)^2},
e^{\delta_{2}}= \frac{(k_{2}-k_{1})^2}{(k_{2}+k_{2}^*)^2(k_{1}+k_{2}^*)^2},
$$ $$
e^{R_{3}}=\frac{|(k_{1}-k_{2})|^4}{(k_{1}+k_{1}^*)^2(k_{2}+k_{2}^*)^2 (|(k_{1}+k_{2}^*|)^4}
.$$

\end{document}